\documentstyle[12pt,aaspp4]{article}
\makeatletter
\let\@internalcite\cite
\def\cite{\def\@citeseppen{-1000}%
    \def\@cite##1##2{(##1\if@tempswa , ##2\fi)}%
    \def\citeauthoryear##1##2##3{##1 ##3}\@internalcite}

\def\citeNP{\def\@citeseppen{-1000}%
    \def\@cite##1##2{##1\if@tempswa , ##2\fi}%
    \def\citeauthoryear##1##2##3{##1 ##3}\@internalcite}
\def\citeN{\def\@citeseppen{-1000}%
    \def\@cite##1##2{##1\if@tempswa , ##2)\else{)}\fi}%
    \def\citeauthoryear##1##2##3{##1 (##3}\@citedata}
\def\citeA{\def\@citeseppen{-1000}%
    \def\@cite##1##2{(##1\if@tempswa , ##2\fi)}%
    \def\citeauthoryear##1##2##3{##1}\@internalcite}
\def\citeANP{\def\@citeseppen{-1000}%
    \def\@cite##1##2{##1\if@tempswa , ##2\fi}%
    \def\citeauthoryear##1##2##3{##1}\@internalcite}
\def\shortcite{\def\@citeseppen{-1000}%
    \def\@cite##1##2{(##1\if@tempswa , ##2\fi)}%
    \def\citeauthoryear##1##2##3{##2 ##3}\@internalcite}
\def\shortciteNP{\def\@citeseppen{-1000}%
    \def\@cite##1##2{##1\if@tempswa , ##2\fi}%
    \def\citeauthoryear##1##2##3{##2 ##3}\@internalcite}
\def\shortciteN{\def\@citeseppen{-1000}%
    \def\@cite##1##2{##1\if@tempswa , ##2)\else{)}\fi}%
    \def\citeauthoryear##1##2##3{##2 (##3}\@citedata}
\def\shortciteA{\def\@citeseppen{-1000}%
    \def\@cite##1##2{(##1\if@tempswa , ##2\fi)}%
    \def\citeauthoryear##1##2##3{##2}\@internalcite}
\def\shortciteANP{\def\@citeseppen{-1000}%
    \def\@cite##1##2{##1\if@tempswa , ##2\fi}%
    \def\citeauthoryear##1##2##3{##2}\@internalcite}
\def\citeyear{\def\@citeseppen{-1000}%
    \def\@cite##1##2{(##1\if@tempswa , ##2\fi)}%
    \def\citeauthoryear##1##2##3{##3}\@citedata}
\def\citeyearNP{\def\@citeseppen{-1000}%
    \def\@cite##1##2{##1\if@tempswa , ##2\fi}%
    \def\citeauthoryear##1##2##3{##3}\@citedata}

\def\@citedata{%
	\@ifnextchar [{\@tempswatrue\@citedatax}%
				  {\@tempswafalse\@citedatax[]}%
}

\def\@citedatax[#1]#2{%
\if@filesw\immediate\write\@auxout{\string\citation{#2}}\fi%
  \def\@citea{}\@cite{\@for\@citeb:=#2\do%
    {\@citea\def\@citea{, }\@ifundefined
       {b@\@citeb}{{\bf ?}%
       \@warning{Citation `\@citeb' on page \thepage \space undefined}}%
{\csname b@\@citeb\endcsname}}}{#1}}%

\def\@citex[#1]#2{%
\if@filesw\immediate\write\@auxout{\string\citation{#2}}\fi%
  \def\@citea{}\@cite{\@for\@citeb:=#2\do%
    {\@citea\def\@citea{; }\@ifundefined
       {b@\@citeb}{{\bf ?}%
       \@warning{Citation `\@citeb' on page \thepage \space undefined}}%
{\csname b@\@citeb\endcsname}}}{#1}}%

\def\@biblabel#1{}

\newlength{\bibhang}
\makeatother

\def\cf{cf.}
\def\ebv{$E($\bv)}

\newcommand{\fig}[1]{Fig.~\ref{#1}}

\def\kms{km s$^{-1}$}

\def\cop{{\it Copernicus}}
\def\hone{\ion{H}{1}}
\def\htwo{H$_2$}

\def\nht{$N$(H$_2$)}
\def\orf{ORFEUS}
\def\sag{Sagittarius}
\def\sodium{\ion{Na}{1}}

\begin{document}
 
\title{ORFEUS-I Observations of Molecular Hydrogen\\
in the Galactic Disk\footnotemark}
\footnotetext{Based on the development and utilization of
\orf\ (Orbiting and Retrievable Far and Extreme Ultraviolet
Spectrometers), a collaboration of the Astronomical Institute of the
University of T\"{u}bingen, the Space Astrophysics Group of the
University of California at Berkeley, and the Landessternwarte
Heidelberg.}


\vspace{0.25in}
\author{W. Van Dyke Dixon, Mark Hurwitz, and Stuart Bowyer}
\affil{Space Sciences Laboratory and Center for EUV Astrophysics\\
University of California, Berkeley, California 94720-5030\\
vand, markh, bowyer@ssl.berkeley.edu}


\vspace{0.5in}
\begin{abstract}

\noindent We present measurements of interstellar \htwo\ absorption lines in the
continuum spectra of seven early-type stars in the Galactic disk at
distances between 1 and 4 kpc.  Five of these stars provide lines of
sight through the \sag\ spiral arm.  The spectra, obtained with the
Berkeley EUV/FUV spectrometer on the \orf\ telescope in 1993 September,
have a resolution of 3000 and statistical signal-to-noise ratios
between 20 and 80.  We determine column densities for each observed
rotational level and derive excitation temperatures and densities for
the \htwo\ clouds along each line of sight.  Our data continue the
relationships among \htwo\ column density, fractional molecular
abundance, and reddening apparent in \cop\ observations of nearby
stars, indicating a common mechanism for \htwo\ production. Estimates
of cloud temperatures and densities are consistent with those derived
from \cop\ data. The molecular fraction $f$ is nearly constant over a
wide range of distances and mean reddenings, consistent with a model in
which a significant fraction of the neutral ISM is associated with
\htwo-bearing molecular clouds, even along low-density lines of sight.

\end{abstract}
 

\newpage

\section{INTRODUCTION}

The hydrogen molecule (\htwo) plays a central role in a variety of
processes that significantly influence the chemical and physical state
of the interstellar medium (ISM).  From observations by \cop\ and other
spacecraft-borne observatories, a picture has emerged in which the bulk
of interstellar \htwo\ lies in clouds with densities between $\sim$ 10
and a few 1000 cm$^{-3}$, diameters less than a few tens of parsecs,
and column densities $\gtrsim 10^{20}$ cm$^{-2}$, which allow the rapid
formation of \htwo\ on dust grains and provide self-shielding against
dissociating photons \cite{WW89}.  In order to determine whether this
model, developed from observations of relatively nearby stars ($d
\lesssim 2$ kpc), holds at larger distances, we observed seven stars in
the Galactic disk ($|z| < 300$ pc) at distances out to 3.9 kpc using
the Berkeley spectrometer on the \orf\ telescope.

\section{OBSERVATIONS AND DATA REDUCTION}
\label{reduction}

The Berkeley EUV/FUV spectrometer, located at the prime focus of the
1-m \orf\ telescope, flew aboard the space platform {\it ASTRO-SPAS\/}
during the 1993 September mission of the space shuttle {\it
Discovery}.  The \orf\ project and the {\it ASTRO-SPAS\/} platform are
described in \citeN{Getal91p437}, while the design and performance of
the Berkeley spectrometer are discussed in \citeANP{HB91p442}
\citeyear{HB91p442,HB95} and \citeN{HBKL95}.

\begin{figure}
\plotone{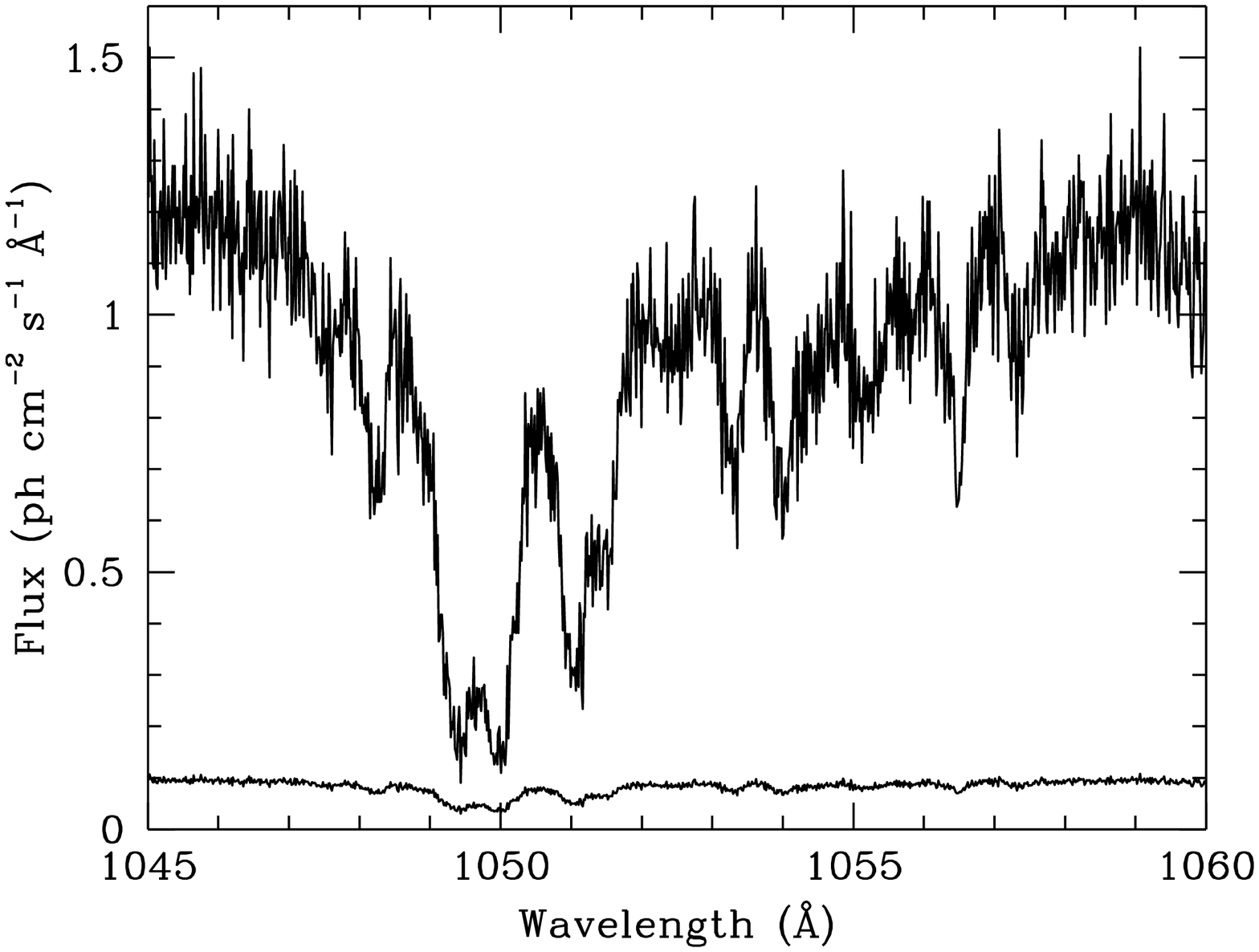}
\figcaption{A portion of the \orf\ spectrum of HD 99890. The spectrum has
been background subtracted and flux calibrated and is plotted in photon
units at full resolution. Principal absorption features are identified
in \fig{model}.  The error spectrum is also shown.
\label{data}
}
\end{figure}

With an effective area of about 4 cm$^2$ and a resolution
$\lambda/\Delta \lambda = 3000$ over the wavelength range 390--1170
\AA, the Berkeley spectrometer is ideally suited for absorption-line
studies of bright, far-UV sources.  Some two dozen early-type stars
were observed on the  \orf-I mission; in this paper, we discuss the
seven with $|z| < 300$ pc. Table \ref{targets} lists the stars and
their observational parameters.  Also given are the total
\orf\ integration time and the statistical signal-to-noise ratio (S/N) 
in a 0.2 \AA\ bin averaged over the 1045--1060 \AA\ band in the
resulting spectrum.  The procedures for data extraction, background
subtraction, and wavelength and flux calibration are described in
\citeN{HB96}.  The flux calibration is based on in-flight observations
of the hot white dwarf G191-B2B and is thought to be uncertain to less
than 10\% \cite{VF92}.  A segment of the \orf\ spectrum of HD 99890 is
presented in \fig{data}; the spectrum is rich in molecular hydrogen
features and the atomic lines of H, N, Ar, and Fe.

We model the interstellar absorption features using an ISM line-fitting
package written by M. Hurwitz and V. Saba. Given the column density,
Doppler broadening parameter, and relative velocity of a given species,
the program computes a Voigt profile for each line, convolves the lines
with a Gaussian to the instrument resolution, and uses the result as a
transmission function by which to scale the model continuum. Given
these assumptions, the only free parameters in the fit are the
continuum placement and the total column density of each absorbing
species.

In principle, it is possible to derive the effective Doppler parameter
directly from the ORFEUS data by measuring the equivalent widths of
many absorption features from a given $J\arcsec$ level, sampling a
range of oscillator strengths.  In practice, line blending, continuum
placement uncertainties, and the limited dynamic range of the
accessible oscillator strengths make such analyses difficult.
The resulting constraints on the effective Doppler parameters
(typically 5--8 \kms) almost certainly reflect velocity-component
structure rather than line broadening in a single cloud. Instead, for
the four lines of sight listed in Table \ref{sembach}, we take the
Doppler parameters, relative velocities, and relative column densities
of the principal absorption components from the literature.  For the
remaining lines of sight, we assume a single absorption component with
a Doppler parameter $b = 5$ km s$^{-1}$, consistent with the $b$ values
of 3--5 \kms\ inferred for the $J\arcsec =$ 2 to 4 lines of \htwo\ from
\cop\ observations (\citeNP{DB86} and references therein). The
$J\arcsec = 0$ and 1 lines lie on the square-root part of the curve of
growth and are thus insensitive to small uncertainties in $b$
\cite{SBDB77}.

To provide an estimate of the stellar continuum, we make use of
reference spectra selected from the 0.2 \AA\ \cop\ Spectral Atlas
\cite{SJ77}. Atlas counterparts, listed in Table \ref{atlas}, are
selected for their similarity to the observed stars in MK
classification and, where possible, in $v \sin i$. The atlas spectra
are smoothed to the resolution of the \orf\ data, re-sampled at the
\orf\ pixel size, and scaled by a linear function to reproduce the
observed stellar continuum. Because the total \htwo\ column density
\nht\ in each reference spectrum is less than the uncertainty in either
the $J\arcsec = 0$ or $J\arcsec = 1$ column of the corresponding
program star, we do not attempt to correct for \htwo\ absorption in the
reference spectrum.  For HD 93129a, an O3 If* star, we expect the
photospheric lines to be quite weak \cite{TKHBPPBLH97} and therefore
assume a linear continuum in modeling the interstellar absorption in
its spectrum.

Following \citeN{HB96}, we adopt 0.31 \AA\ as the FWHM of the
instrument point-spread function.  In the course of this work, we have
discovered small nonlinearities in the \orf\ wavelength calibration
that will have to be quantified before the velocities of the absorbing
clouds can be constrained. For this analysis, we assume that the clouds
have zero radial velocity and adapt the wavelength scale accordingly.

\begin{figure}
\plotone{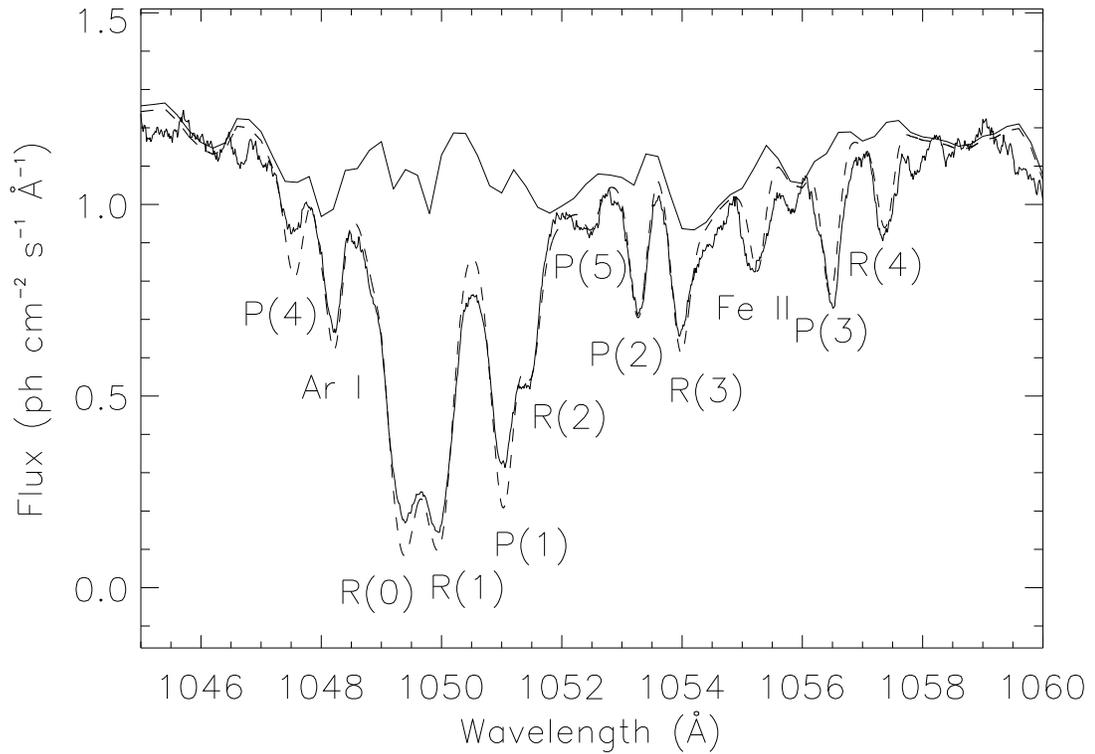}
\figcaption{The $v = 0 \rightarrow 4$ vibrational band in the Lyman series
of molecular hydrogen from the \orf\ spectrum of HD 99890. The data
have been smoothed with a 13-pixel boxcar.  Overplotted are our
best-fitting model (dashed line) and the spectrum of $\theta$ Car from
the \cop\ Spectral Atlas (Snow \& Jenkins 1977), which we use to model
the stellar features in the HD 99890 spectrum.
\label{model}
}
\end{figure}

To perform the model fit, we concentrate on the region between 1045 and
1060 \AA, which contains most of the $v = 0 \rightarrow 4$ vibrational
band in the Lyman series, as this region seems least complicated by
stellar photospheric absorption or overlapping Werner bands of
interstellar \htwo. The process is iterative:  we begin by fitting the
$J\arcsec = 0$ and $J\arcsec = 1$ lines, move on to the
higher-$J\arcsec$ lines, then repeat the process, as the
higher-$J\arcsec$ lines often lie on the wings of lower-$J\arcsec$
features. Isolated features elsewhere in the spectrum provide a
double-check on our results. The large number of free parameters and
the computationally-intensive nature of our fitting routine make an
automated $\chi^2$-reduction algorithm impractical. Instead, we vary
the model column density about the best-fit value until the model line
profile departs significantly from the data and use this as an estimate
of the uncertainty in the fit. Column densities derived in this way for
the $J\arcsec = 0$ to 5 rotational levels of \htwo\ are presented in
Table \ref{results}. We estimate uncertainties of 0.2 dex for $J\arcsec
= 0$ to 3 and 0.5 dex for $J\arcsec \geq 4$.  \fig{model} presents the
spectrum of HD 99890 between 1045 and 1060 \AA\ together with our
best-fitting model.

To investigate the uncertainty introduced by assuming a single
absorption component in the spectra of stars lacking published
absorption-line parameters, we generate a series of synthetic spectra
(with S/N $\sim$ 35), each with two absorption components of equal
column density separated by 5, 15, 25, 35, and 45 \kms, respectively.
To each synthetic spectrum, we fit a model assuming a single velocity
component.  For the $J\arcsec = 0$ and $J\arcsec = 1$ states, the
derived column densities differ by less than our quoted uncertainty
over the entire 5--45 \kms\ range.  For the $J\arcsec = 2$ and
$J\arcsec = 3$ states, however, the difference can be significant: as
great as 1.5 dex for the $J\arcsec = 3$ state in the 25 and 35
\kms\ models, in the sense that a single-cloud model over-estimates the
total column density in the $J\arcsec = 3$ state.  The $J\arcsec = 4$
and $J\arcsec = 5$ states vary, but not by more than our estimated
uncertainties (0.5 dex).  Fortunately, the total column \nht\ is
dominated by the $J\arcsec = 0$ and $J\arcsec = 1$ states \cite{SCH74}.
The values of $n$ and $T_{01}$ derived in \S~\ref{params} are thus
insensitive to uncertainties in the velocity structure of the absorbing
medium.

\section{ANALYSIS}
\label{params}

The strongest UV lines arise from the $J\arcsec = 0$ and $J\arcsec = 1$
levels, whose relative populations are established by thermal proton
collisions, since the molecules are protected from photo-excitation by
self-shielding in the lines. The mean excitation temperature of the
clouds along each line of sight can thus be derived from the column
densities $N(0)$ and $N(1)$ using the relation
\begin{equation}
\frac{N(1)}{N(0)} = \frac{g_1}{g_0} \exp \left( \frac{-E_{01}}{k T_{01}} \right)
= 9 \exp \left( \frac{-170 \; {\rm K}}{T_{01}} \right),
\end{equation}
where $g_0$ and $g_1$ are the statistical weights of $J\arcsec = 0$ and
$J\arcsec = 1$, respectively \cite{SB82}.  Cloud
temperatures for the \orf\ lines of sight are presented in Table
\ref{results}.  Our results are consistent with those obtained with
\cop\ by \citeANP{SBDB77}, who derive values of 
$T_{01}$ ranging from 45 to 128 K, with an average of $77 \pm 17$ (rms) K,
for 61 stars with \nht\ $> 10^{18}$ cm$^{-2}$.

We estimate the cloud density $n$ using the relation derived by
\citeN{RKH94},
\begin{equation}
\Biggl( \frac{n}{50 \; {\rm cm}^{-3}} \Biggr)^2 \simeq
\Biggl[ \frac {N({\rm H}_2)}{10^{19} \; {\rm cm}^{-2}} \Biggr] \;
\Biggl( \frac{T}{80\; {\rm K}} \Biggr)^{-1} \;
\Biggl[ \frac{N({\rm H\;I})}{10^{20} \; {\rm cm}^{-2}} \Biggr]^{-2} ,
\end{equation}
where $n$ is the proton (\hone\ + 2\htwo) volume density and the values
of $N$(\hone), \nht, and $T_{01}$ are from Table \ref{results}.  The
resulting cloud densities, presented in Table \ref{results}, are lower
limits for two reasons: first, because an unknown fraction of
$N$(\hone) lies outside of the \htwo-bearing clouds and, second, because
the relation assumes a single cloud along the line of sight.  If the
observed \hone\ and \htwo\ columns were divided evenly among $n_c$
clouds, for example, then the density of each cloud would rise
as $\sqrt{n_c}$.  The density limits derived via equation (2) are 
consistent with the values between 15 and 900 cm$^{-3}$ found by
\citeN{Jura75} for interstellar clouds containing optically thick
\htwo.

While detailed modeling of the relative $J\arcsec$ column densities is
beyond the scope of this paper, we can investigate the radiation
environments of the \htwo\ clouds by comparing our observations with
the predictions of previously-published models. \citeN{Jura75} models
the $J\arcsec =$ 0 to 6 column densities along ten lines of sight
observed with \cop.  He finds that the ratios $N$(4)/$N$(0) and
$N$(5)/$N$(1) are controlled by the optical pumping rate, rather than
by the electron temperature of the cloud, and concludes that clouds
along four lines of sight are illuminated by an ultraviolet field
several times that in the solar neighborhood. All four of these sight
lines show values of $\log [N(4)/N(0)] > -4.8$. (A fifth sight line
also showing a high value of $N$(4)/$N$(0) can be fit without recourse
to an elevated radiation field.) In Table \ref{results}, we see that
five of our seven stars have $\log [N(4)/N(0)] > -4.8$. Given that the
uncertainty in this ratio is $\sim$0.5 dex, we consider only the
three stars with the highest values of $N(4)/N(0)$, HD~93129a,
HD~99857, and HD~99890. While his models assume a single absorbing cloud,
\citeANP{Jura75} points out that, if there are several clouds along the
line of sight, the ultraviolet opacity within each cloud is lower than
if only one cloud is present, and the observations can often be fit
with several clouds at normal UV levels, rather than a single cloud
with elevated illumination. Multiple-cloud models may be able to
reproduce the high $N(4)/N(0)$ ratios that we observe. On the other
hand, the absorbing clouds may simply lie near the hot background
stars. For HD~93129a in the Carina Nebula, this is certainly the case
(cf., \citeNP{WHH84}).

\section{DISCUSSION}

\begin{figure}
\plotone{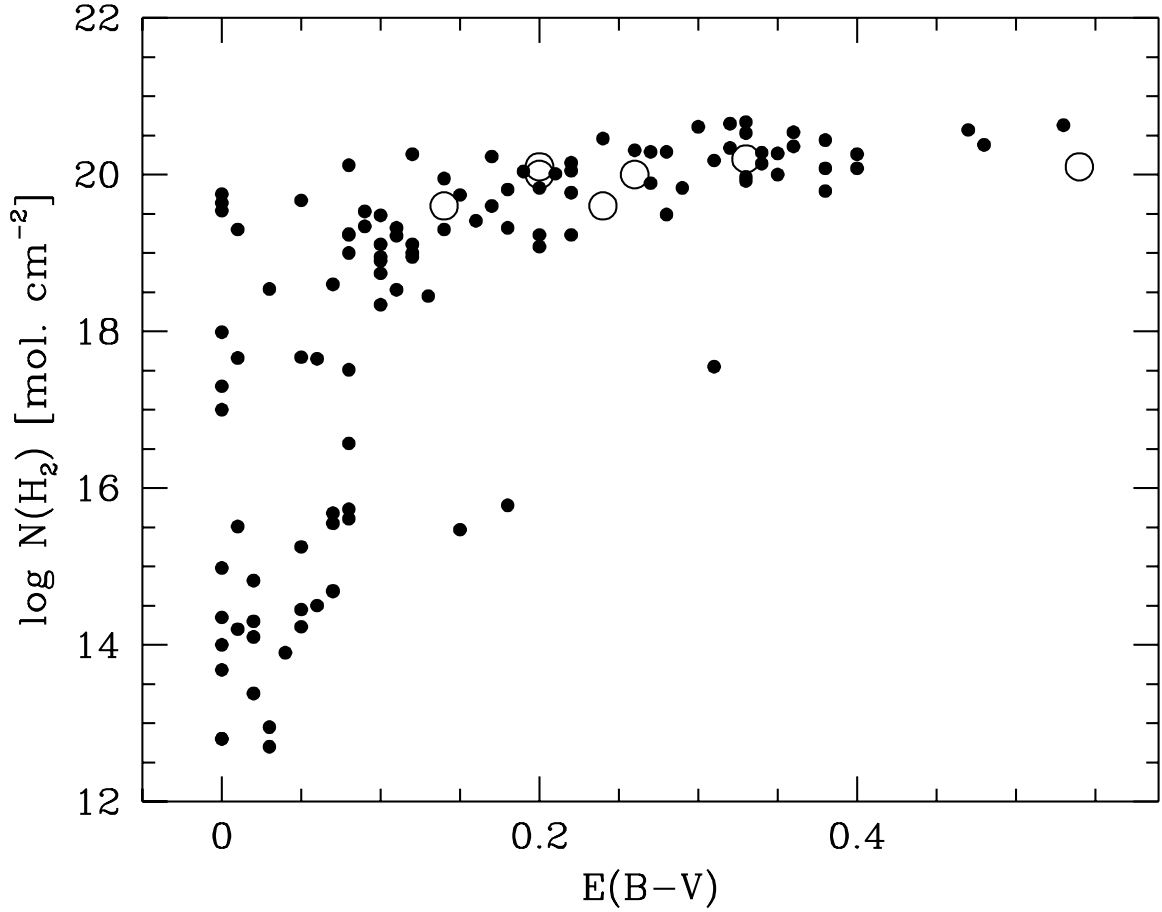}
\figcaption{The logarithm of the total \htwo\ column density versus \ebv,
a measure of the dust column density. Solid symbols 
represent lines of sight observed with \cop\ (Savage et al. 1977).
Many of the \cop\ points with \ebv\ $< 0.1$ are upper limits.
Open symbols represent the seven stars observed with \orf.
\label{nh2_ebv}
}
\end{figure}

\begin{figure}
\plotone{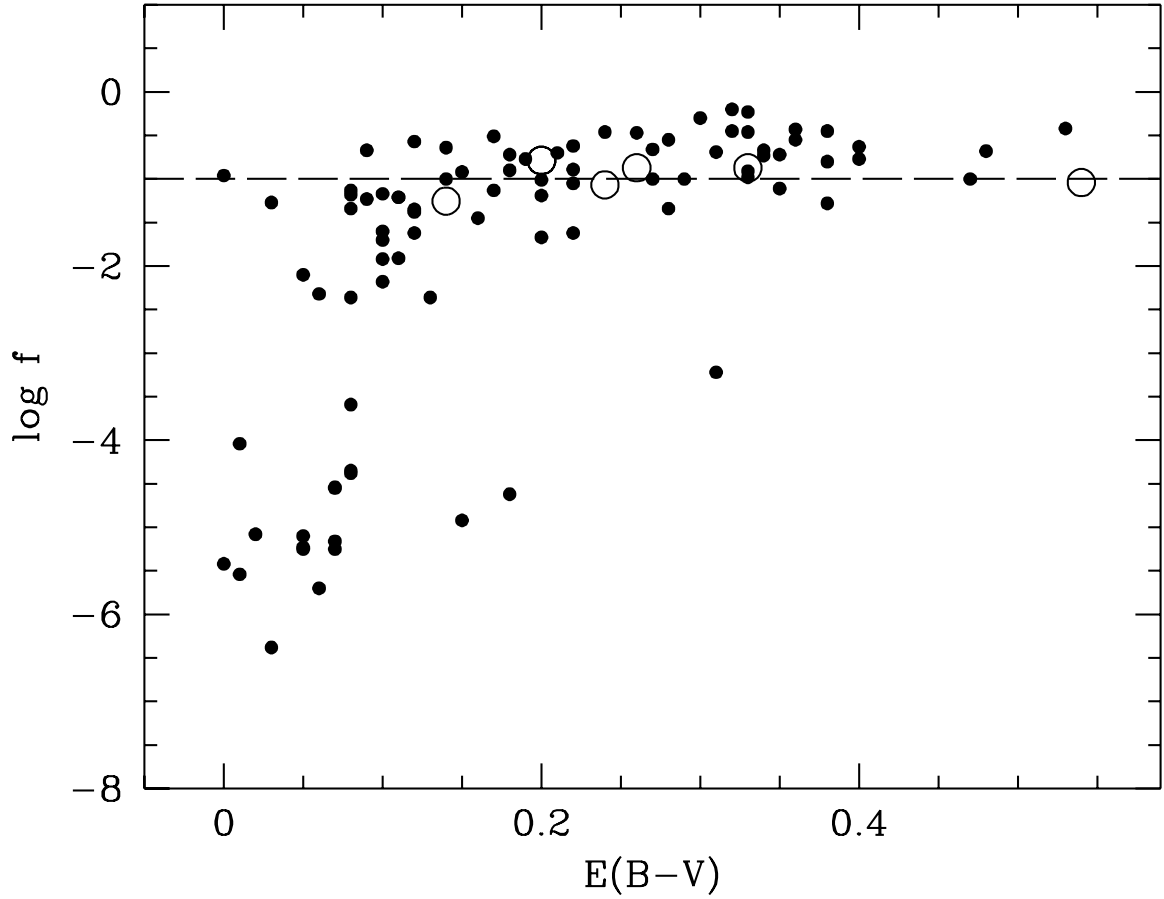}
\figcaption{The logarithm of $f$, the fractional abundance of \htwo,
versus \ebv. A dashed line marks the level at which $\log f = -1.0$.
Solid symbols represent \cop\ observations, while the open symbols
represent this work. Two \orf\ data points with \ebv\ = 0.20 overlap.
\label{logf_ebv}
}
\end{figure}

\begin{figure}
\plotone{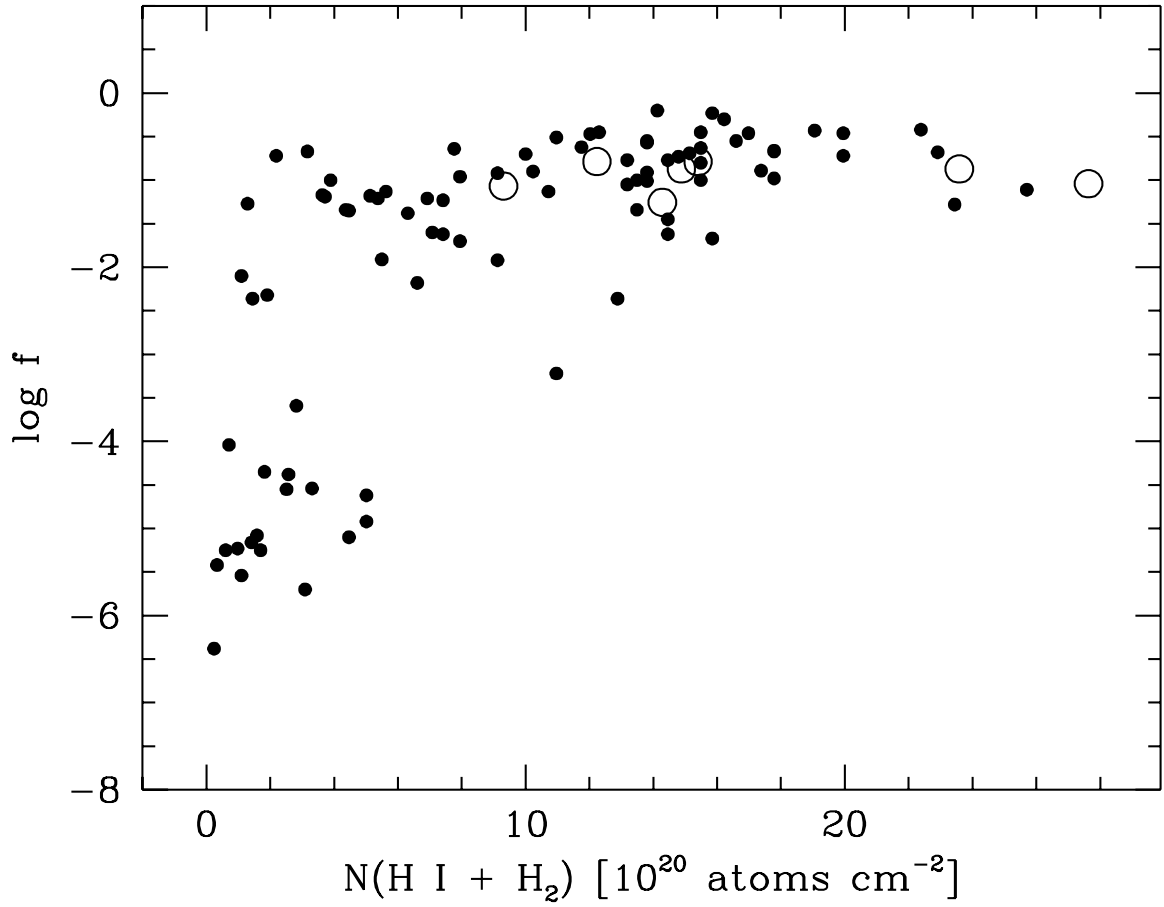}
\figcaption{A plot of $\log f$ versus the total hydrogen column density,
$N($\hone$ + {\rm H}_2) = N($\hone$) + 2N({\rm H}_2)$.  Solid symbols
represent \cop\ observations, while the open symbols are from this
work.
\label{logf_nh12}
}
\end{figure}

Figures \ref{nh2_ebv}, \ref{logf_ebv}, and \ref{logf_nh12}, based on
Figures 4--6 of \citeN{SBDB77}, explore the relationships among the
\htwo\ column density \nht, the fraction of hydrogen in the molecular
state $f = 2N($H$_2) / [N($\hone$) + 2N($H$_2)]$, the reddening \ebv,
and the total hydrogen column density $N($\hone$ + {\rm H}_2) =
N($\hone$) + 2N({\rm H}_2)$.  (Many of the \cop\ data points in the
lower left of each diagram are actually upper limits.) \citeANP{SBDB77}
found that lines of sight with low extinction [\ebv $\lesssim 0.08$]
show low columns of \htwo\ and low values of $f$ ($\sim 0.02$), while
those with higher extinction [\ebv $> 0.08]$ show much higher \htwo\
columns and values of $f$ that rise slowly with \ebv.  The change from
small to large $f$ at \ebv\ $\approx 0.08$ and $N($\hone$ + $\htwo$)
\approx 5 \times 10^{20}$ cm$^{-2}$ is thought to reflect the column
density at which the diffuse \htwo-bearing clouds provide
self-shielding against radiative dissociation.

The seven \orf\ disk stars fall squarely in the ``large~$f$'' region of
all three figures, indicating a common mechanism for
\htwo\ production.  We have seen that the cloud temperature and density
limits derived from the \orf\ observations are consistent with those
obtained with \cop.  From the \htwo\ column densities $N({\rm H}) = 1-3
\times 10^{21}$ cm$^{-2}$ and proton densities $n \gtrsim 10$ cm$^{-3}$
in Table \ref{results}, we estimate path lengths on the order of tens
of parsecs, consistent with standard models \cite{WW89}.

The five stars in our sample with $d > 3$ kpc all lie at galactic
longitudes between 287\arcdeg\ and 298\arcdeg, providing lines of sight
directly through the Sagittarius spiral arm.  (The distances, taken
from the literature, were originally derived from stellar spectral
types and can be assumed accurate to about 25\%; \cf, \citeNP{SDS93}.)
Unless the Sagittarius arm is devoid of molecular gas, our results
suggest that its clouds do not differ significantly in their gross
properties from those near the sun.

\citeN{Sembach94} has recently performed a detailed study of highly
ionized gas along this line of sight $(l = 295\arcdeg, \; b =
0\arcdeg)$ and finds that it samples mainly low-density disk gas in the
form of inter-cloud and diffuse cloud material. \citeN{SD94} find that
the \sodium\ and \ion{Ca}{2} absorption profiles along such low-density
sight lines show considerable structure from diffuse clouds. The number
of components in the profiles is loosely correlated with $n_0$(\hone),
the average density of \hone\ in the Galactic disk derived for a given
line of sight \shortcite{SDS93}, suggestive of a patchy
distribution for the neutral ISM along low-density lines of sight
\cite{SD94}.

\citeN{SBDB77} point out that \cop\ sampled lines of sight with
relatively low reddening. While the mean reddening $[\Sigma
E(B-V)/\Sigma r]$ in the Galactic plane within 1 kpc of the Sun is
0.61 mag kpc$^{-1}$ \cite{Spitzer68p67}, the mean reddening to the
\cop\ stars is 0.22 mag kpc$^{-1}$. The mean reddening to the
\orf\ disk stars is even lower, 0.11 mag kpc$^{-1}$.

\begin{figure}
\plotone{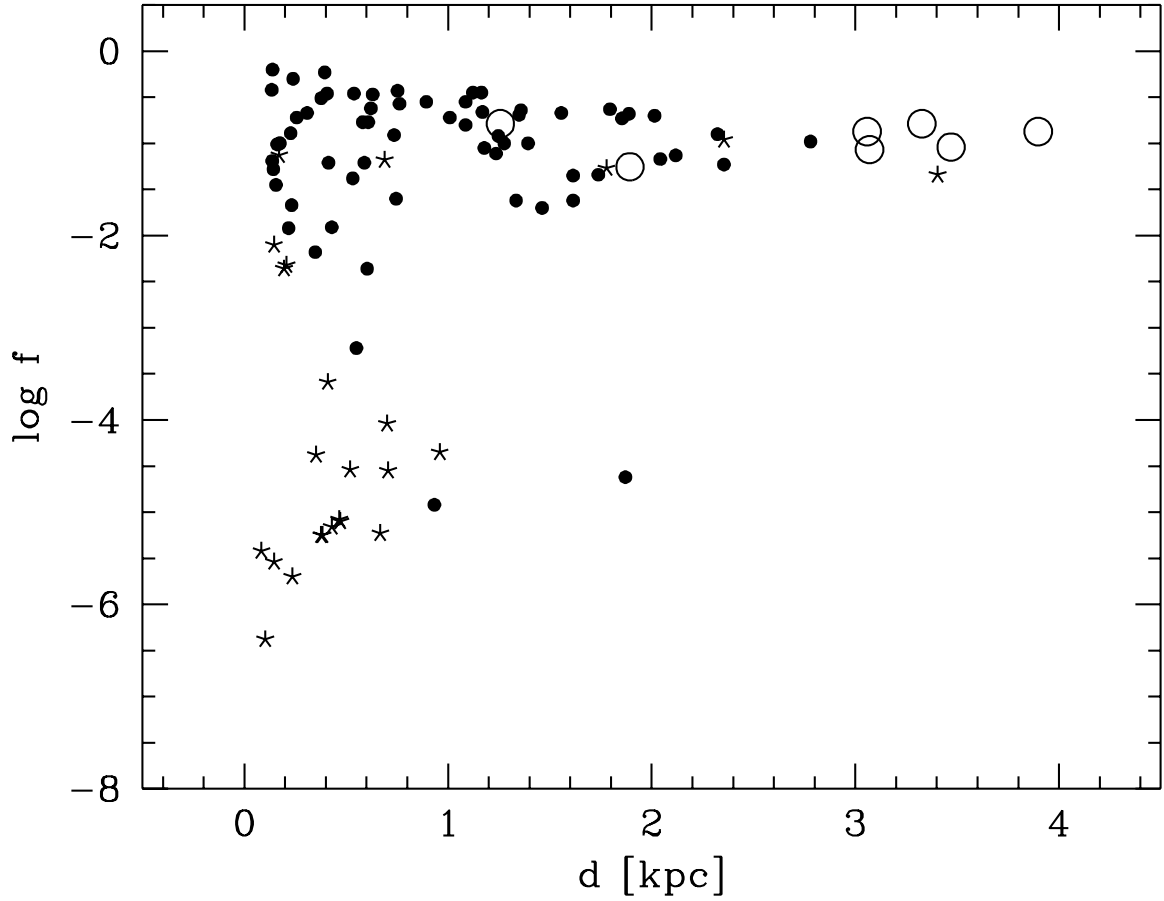}
\figcaption{A plot of $\log f$ versus the derived stellar distance.  Solid
symbols represent \cop\ observations, except for stars with \ebv\ $\leq
0.08$, which are plotted here as asterisks. The open symbols are from
this work. We find that $f$ remains $\sim$ 10\% even at large distances
in the disk.
}
\label{logf_d}
\end{figure}

In \fig{logf_d}, we plot $\log f$ versus stellar distance. To highlight
the different behaviors along high- and low-reddening lines of sight,
stars with $E(B-V) \leq 0.08$ are plotted as asterisks.  \citeN{SBDB77}
find a slow increase in $f$ with \ebv\ for $E(B-V) > 0.15$, a trend
confirmed in our data (\fig{logf_ebv}).  As a function of distance,
however, we see that $f$ remains nearly flat out to 4 kpc.  Because our
sight lines do not sample Galactic radii very different from that of
the sun, we do not expect $f$ to approach unity at large distances, as
it does toward the inner Galaxy \cite{Dame93p267}.  Nevertheless, it is
remarkable that the molecular fraction of hydrogen should vary so
little over the wide range of distances and mean densities sampled by
\cop\ and \orf. If the neutral and molecular hydrogen were not coupled,
one might expect significant variations in the molecular fraction of
hydrogen.  On the other hand, a constant value of $f$ is predicted by
models in which most of the neutral ISM is associated with the
\htwo-bearing clouds, perhaps in a shell surrounding the molecular core
(\cf, \citeNP{MO77}).

\section{CONCLUSIONS}

We have measured column densities for the $J\arcsec = 0$ to 5
rotational states of \htwo\ in the spectra of seven disk stars ($|z| <
300$ pc) observed with the Berkeley EUV/FUV spectrometer on the \orf\
telescope. These stars lie at distances of up to 3.9 kpc, allowing us
to probe lines of sight much deeper into the Galactic disk than were
heretofore available.  Five of our sight lines intersect the \sag\
spiral arm.  Our data continue the relationship between \htwo\ column
density and reddening established for the \cop\ clouds, indicating a
common mechanism for \htwo\ production. Estimates of cloud temperatures
and densities are consistent with those derived from
\cop\ observations. The molecular fraction $f$ is nearly constant over
a wide range of distances and mean reddenings, consistent with a model
in which the neutral component of the ISM is associated mainly with the
\htwo-bearing clouds.

\acknowledgments

This research has made use of the NASA ADS Abstract Service and the
Catalogue Service of the CDS, Strasbourg, France. We thank J. Black for
providing \htwo\ transition data in electronic format and C. McKee for
helpful comments on the text.  We acknowledge our colleagues on the
\orf\ team and the many NASA and DARA personnel who helped make the
\orf-I mission successful. This work is supported by NASA grant
NAG5-696.


\clearpage 

\begin{deluxetable}{ccclcccccc}
\tablewidth{0pt}
\tablecaption{Target Summary\label{targets}}
\tablehead{
\multicolumn{5}{c}{} &
\colhead{$d$}	& \colhead{$v \sin i$} &
\colhead{$T$} \\
\colhead{HD}   &
\colhead{\it l}         & \colhead{\it b}  &
\colhead{Sp. Type}	& \colhead{$E(B-V)$}  &
\colhead{(pc)}	& \colhead{(km s$^{-1}$)}  &
\colhead{(s)}	& \colhead{S/N}  &
\colhead{References}
}
\startdata
\phn41161\phm{a} & 165.0 &    $+$12.9 & O8 V        & 0.20 & 1253 & 300\phm{:} &
\phn814 & 45.2 & 1,2 \nl
\phn54911\phm{a} & 229.0 & \phn$-$3.1 & B1 III      & 0.14 & 1893 & 100: &
2462 & 78.0 & 1,3 \nl
\phn93129a       & 287.4 & \phn$-$0.6 & O3 If*      & 0.54 & 3470 & 120\phm{:} &
3045 & 36.2 & 4,5 \nl
\phn94493\phm{a} & 289.0 & \phn$-$1.2 & B0.5 Iab/Ib & 0.20 & 3327 & 145\phm{:} &
1528 & 34.9 & 1,3 \nl
\phn99857\phm{a} & 294.8 & \phn$-$4.9 & B1 Ib       & 0.33 & 3058 & 180\phm{:} &
1383 & 22.3 & 1,3 \nl
\phn99890\phm{a} & 291.8 & \phn$+$4.4 & B0.5 V:     & 0.24 & 3070 & 180\phm{:} &
2532 & 39.7 & 1,3 \nl
104705\phm{a}    & 297.5 & \phn$-$0.3 & B0 III/IV   & 0.26 & 3898 & 215\phm{:} &
1788 & 36.3 & 1,3
\enddata
\tablerefs{(1) Fruscione et al. 1994;
(2) rotational velocity from Jenkins 1978;
(3) rotational velocity from Savage \& Massa 1987;
(4) Walborn 1973; (5) Gies 1987.}
\end{deluxetable}
\nocite{FHJW94,Gies87,SM87,Jenkins78,Walborn73}

\begin{deluxetable}{clclllc}
\tablewidth{0pt}
\tablecaption{Measured Parameters of Individual Velocity Components\label{sembach}}
\tablehead{
\multicolumn{2}{c}{}	& \colhead{Component} &
\colhead{$\langle v_i \rangle$}	&
\colhead{$b_i$}		& \colhead{N$_i$} 	\\
\colhead{HD}		& \colhead{Species} &
\colhead{Number}	& \colhead{(km s$^{-1}$)}	&
\colhead{(km s$^{-1}$)}	& \colhead{(cm$^{-2}$)}	&
\colhead{Notes}
}
\startdata
\phn54911\phm{a} & \sodium\ & 1 & \phm{$-$1}7.5$\pm$0.1 & 3.0$\pm$0.1 & 3.90$\pm$0.18(12) & 1 \nl
		 &          & 2 & \phm{$-$}18.3$\pm$0.1 & 1.8$\pm$0.1 & 1.71$\pm$0.12(12) &   \nl
\phn93129a       & \ion{Ca}{2}, \sodium\ & 1 & $-33$    & \nodata     & \nodata           & 2 \nl
		 &          & 2 & \phm{$-$1}0           & \nodata     & \nodata           &   \nl
\phn94493\phm{a} & \sodium\ & 1 & \phm{$-$}17.7$\pm$0.1 & 5.6$\pm$0.1 & 5.52$\pm$0.07(12) & 1 \nl
		 &          & 2 & \phm{$-$1}0.6$\pm$0.1 & 4.0$\pm$0.1 & 3.55$\pm$0.07(12) &   \nl
104705\phm{a}    & \sodium\ & 1 & $-$30.7$\pm$0.2       & 2.2$\pm$0.2 & 1.69$\pm$0.46(12) & 1 \nl
		 &          & 2 & $-$22.2$\pm$0.2       & 2.4$\pm$0.2 & 1.23$\pm$0.13(12) &   \nl
		 &          & 3 & \phm{$-$1}0.3$\pm$0.1 & 3.4$\pm$0.1 & 5.27$\pm$0.39(12) &
\enddata
\tablecomments{(1) Principal absorption components along lines of sight
observed by Sembach et al. 1993;
(2) Walborn 1982 find a number of weak interstellar features between $-13$
and $+19$ km s$^{-1}$, which we render as a single velocity component at 0
km s$^{-1}$ with a column density twice that of the $-33$ km s$^{-1}$ 
feature. We assume $b = 5$ km s$^{-1}$ for both components.}
\end{deluxetable}
\nocite{SDS93,Walborn82}

\begin{deluxetable}{clclccc}
\tablewidth{0pt}
\tablecaption{\cop\/\tablenotemark{a} $\;$ Continuum Reference Stars\label{atlas}}
\tablehead{
\colhead{Program Star}   & \multicolumn{2}{c}{Atlas Star} &
\multicolumn{2}{c}{} & \colhead{$v \sin i$} &
\colhead{$\log$ \nht\tablenotemark{b}} \\ \cline{2-3}
\colhead{HD}   & \colhead{Name}   & \colhead{HD}    &
\colhead{Sp. Type}	& \colhead{$E(B-V)$}  &
\colhead{(km s$^{-1}$)}  & \colhead{(cm$^{-2}$)}
}
\startdata
\phn41161\phm{a} & 15 Mon & \phn47839 & O7 V((f)) & 0.07 & 106 & \phm{$<$}15.55 \nl
\phn54911\phm{a} & $\beta$ Cen A & 122451 & B1 III & 0.02 & \phn70 & \phm{$<$}12.8\phn \nl
\phn93129a       & None \nl
\phn94493\phm{a} & $\kappa$ Ori & \phn38771 & B0.5 Ia & 0.07 & \phn81 & \phm{$<$}15.68 \nl
\phn99857\phm{a} & $\kappa$ Ori \nl
\phn99890\phm{a} & $\theta$ Car & \phn93030 & B0.5 Vp & 0.06 & 130 & $<$17.65 \nl
104705\phm{a}    & $\kappa$ Ori \nl
\enddata
\tablenotetext{a}{Snow \& Jenkins 1977.}
\tablenotetext{b}{Savage et al. 1977.}
\end{deluxetable}
\nocite{SJ77,SBDB77}

\begin{deluxetable}{lccccccccccc}
\small
\tablewidth{0pt}
\tablecaption{Column Densities\label{results}}
\tablehead{
&&&&&&&&&& \colhead{$T_{01}$\tablenotemark{c}} &
\colhead{$n$\tablenotemark{d}} \nl
\colhead{HD}   &
\colhead{$N$(\hone)\tablenotemark{a}}         & \colhead{$N$(H$_2$)}  &
\colhead{$N$(0)}	& \colhead{$N$(1)}   &
\colhead{$N$(2)}	& \colhead{$N$(3)}   &
\colhead{$N$(4)}	& \colhead{$N$(5)}   &
\colhead{{$\frac{\mbox{\normalsize N(4)}}
{\mbox{\normalsize N(0)}}$}\tablenotemark{b}}   &
\colhead{(K)}   & \colhead{(cm$^{-3}$)}
}
\startdata
\phn41161\phm{a} & 21.01 & 20.0 & 19.7 & 19.7 & 17.6 & 17.6 & 15.0 & 14.2 & $-4.7$ & 77 & 16 \nl
\phn54911\phm{a} & 21.13 & 19.6 & 19.3 & 19.3 & 17.4 & 16.9 & 14.5 & 13.9 & $-4.8$ & 77 & \phn8 \nl
\phn93129a & 21.40 & 20.1 & 19.7 & 19.9 & 17.2 & 16.9 & 15.9 & 16.1 & $-3.8$ & 98 & \phn6 \nl
\phn94493\phm{a} & 21.11 & 20.1 & 19.7 & 19.8 & 17.7 & 16.2 & 15.0 & 15.1 & $-4.7$ & 86 & 13 \nl
\phn99857\phm{a} & 21.31 & 20.2 & 19.8 & 20.0 & 18.5 & 18.2 & 16.2 & 15.4 & $-3.6$ & 97 & \phn9 \nl
\phn99890\phm{a} & 20.93 & 19.6 & 19.2 & 19.3 & 17.7 & 17.6 & 16.0 & 14.7 & $-3.2$ & 86 & 11 \nl
104705\phm{a} & 21.11 & 20.0 & 19.7 & 19.7 & 17.5 & 16.6 & 14.6 & \nodata & $-5.1$ & 77 & 13 \nl
\enddata
\tablecomments{All column densities are given as logarithms.
Units are cm$^{-2}$. Estimated uncertainties are 0.2 dex for $J\arcsec
= 0$ to 3 and 0.5 dex for $J\arcsec \geq 4$.}
\tablenotetext{a}{\hone\ column densities from the compilation of Fruscione et al. 1994,
except for HD 93129a, which is from Taresch et al. 1997.}
\tablenotetext{b}{$\log [N(4)/N(0)]$. Values greater than $-4.8$ may indicate
the presence of a radiation field several times that in the solar neighborhood.
Uncertainties in this ratio are $\sim 0.5$ dex. See \S~\ref{params}.}
\tablenotetext{c}{Cloud temperature derived from $N$(1)/$N$(0).
Uncertainties are $\sim 33$\%.  See \S~\ref{params}.}
\tablenotetext{d}{Lower limit to cloud proton (\hone\ + 2\htwo) density derived
from $N$(\hone), \nht, and $T_{01}$. See \S~\ref{params}.}
\end{deluxetable}
\nocite{FHJW94,TKHBPPBLH97}

\clearpage 




\newpage 

\nocite{SJ77}

\end{document}